# Frequency and phase noise of ultra-high Q silicon nitride nanomechanical resonators

King Y. Fong, Wolfram H. P. Pernice[*], and Hong X. Tang[†]

*Departments of Electrical Engineering, Yale University,*

*New Haven, CT 06511, USA*

**Abstract**

We describe the measurement and modeling of amplitude noise and phase noise in ultra-high Q nanomechanical resonators made from stoichiometric silicon nitride. With quality factors exceeding 2 million, the resonators' noise performance is studied with high precision. We find that the amplitude noise can be well described by the thermomechanical model, however, the resonators exhibit sizable extra phase noise due to their intrinsic frequency fluctuations. We develop a method to extract the resonator frequency fluctuation of a driven resonator and obtain a noise spectrum with $k_B T / f$ dependence, which could be attributed to defect motion with broadly distributed relaxation times.

PACS numbers: 85.85.+j, 62.25.Fg, 05.40.-a, 63.50.Lm

---

[*] Present address: Institute of Nanotechnology, Karlsruhe Institute of Technology, 76133 Karlsruhe, Germany
[†] Corresponding author: hong.tang@yale.edu

Nanoelectromechanical systems (NEMS) have shown great promise for both fundamental science and applications, by introducing new tools for studying quantum physics [1-3] and enabling high performance devices for sensing and oscillator applications [4-6]. Recently, nanomechanical resonators made of high stress stoichiometric silicon nitride have provided a solution to one of the most challenging quests in NEMS research, i.e., achieving very high mechanical quality factors (Q factors) [7-9]. For a fixed resonance frequency, high Q is equivalent to a low dissipation rate, which implies weaker coupling to the thermal bath, higher sensitivity to external signals, and lower minimum operating power [10]. To enable real world applications based on such devices, understanding the underlying dissipation and noise mechanism is crucial. Such an understanding is also a key for implementing quantum measurements of phonon shot noise as recently proposed [11]. Several models have been developed to explain the damping mechanism [12-14]. Yet, the noise characteristics such as the amplitude/phase noise and frequency fluctuation have not been addressed so far.

In NEMS, resonator frequency fluctuations can arise due to various processes, such as temperature fluctuations [15], adsorption-desorption of molecules on the device surface [16,17], molecule diffusion along the resonator [18]. These frequency fluctuations are usually accompanied by thermal motion of the resonator, which has to be separated to reveal true resonator frequency fluctuations as recently suggested [19]. Here, we experimentally study the frequency fluctuation of ultra high Q nanomechanical resonators made of stoichiometric silicon nitride. We derive and validate a mathematical model to account for the observed frequency fluctuations. Our results show that high Q devices are actually more susceptible to frequency flickering. By measuring the phase noise of the device, a resonator frequency noise spectrum with $k_B T / f$ dependence is extracted. We propose that this fluctuation is due to defect motion with a broad spectrum of thermally-activated relaxation times. Our results give insight into the noise mechanism of high Q silicon nitride resonators and are important for NEMS based applications, such as sensors and low phase noise optomechanical oscillators.

We fabricate nanomechanical resonators from LPCVD stoichiometric silicon nitride. Fig. 1(a) shows the SEM image of a fabricated device, which contains a doubly-clamped beam resonator with dimensions $380\mu m \times 0.33\mu m \times 0.2\mu m$, an optical readout circuit with grating coupler input/output ports as well as an on-chip interferometer. (Fabrication details can be found in Refs. [9,20]). The devices are mounted in a helium cryostat and the temperature of the sample stage is regulated from 5K to room temperature with milli-Kelvin precision. The pressure is kept below $10^{-6}$ Torr for all the measurements carried out. The sample stage is equipped with a piezodisk actuator for mechanical actuation. The mechanical quality factor is measured using a ring-down method [9] and is plotted against temperature in Fig. 1(b). At room-temperature we measure a quality factor of the fundamental in-plane flexural mode of roughly 400,000, which increases up to 2.2 million when the device is cooled down to 5K. The observed high Q of tensile stressed silicon nitride nanostring can be attributed to the increased elastic energy stored in the tension [12,14]. For devices with cross-sectional dimensions similar to the device used here, it was suggested that the Q is limited by internal material damping [12]. It has also been shown that the quality factor of silicon nitride nanostring resonators is not sensitive to pressure change once the pressure is reduced below milli-Torr levels [7,9]. Therefore, we can safely neglect the effect of pressure change on the Q during the cool down.

Under tensile stress, a beam with length of $L$ has a mechanical fundamental mode profile of $\psi(z,t) = x(t)\sin(\pi z/L)$, where the displacement $x(t)$ is defined as the displacement of the middle of the beam. Under such normalization condition, we define the effective mass $m$ as half of the total mass so that $m\dot{x}(t)^2/2$ correctly expresses the total kinetic energy of the resonator. (Details of definition of effective mass can be found, for example, in section 3.1 of Ref [3].) Because of the sensitive optical interferometric read-out, thermomechanical noise is resolved at all temperatures concerned in this work and thus allows us to calibrate the displacement of the resonator. Meanwhile, the optical power is kept minimal at around 10μW in order not to perturb the thermal equilibrium or induce any optomechanical backaction. Device displacement at different driving intensity is plotted in Fig. 1(c). At large drive the device response displays Duffing nonlinearity due to stiffening effects in the displaced beam. If the nonlinear equation of motion $\ddot{x}+2\gamma\dot{x}+\omega_r^2 x+\omega_r^2 K_3 x^3 = F\cos(\omega_d t)/m$ and harmonic motion $x = A\cos(\omega_d t+\phi)$ are assumed, where $\omega_r$ is the resonance frequency, $\gamma = \omega_r/2Q$ is the damping rate, $F$ is the external driving force, $\omega_d$ is the driving frequency and $A$ is the amplitude of the displacement, the nonlinear coefficient $K_3$ can be found by fitting a cubic polynomial to the graph $F^2$ vs $A^2$. For the present case, we obtain $K_3 = 6.9\times10^{-9}\text{nm}^{-2}$, which agrees with the theoretical value $K_3 = \pi^2 E/4\sigma L^2$ expected for a tensile stressed beam experiencing additional stress due to large strain [21]. Here, $E$ and $\sigma$ are the Young's modulus and the stress, respectively.

To separate and quantify the amplitude and phase noise, we used a lock-in amplifier (Zurich Instruments HF2LI [22]) to measure the two quadrature components of the device in response to a sinusoidal drive. The lock-in bandwidth is chosen such that within the bandwidth $\Delta\omega_{LI}$ the noise power is dominated by thermomechanical noise, while $\Delta\omega_{LI} \gg \omega_r/Q$ so that most of the thermal noise power is captured. Fig. 2(a) shows the measured two quadrature components in the complex plane in the presence (lower datasets) and absence (upper datasets) of the drive. The data acquisition time for each dataset is kept at 100s, which is much longer than the device ringdown time. The black solid line traces out the polar plot of the driven response when the driving frequency is swept across the resonance. To ensure linear operation, i.e., $K_3 A^2 \ll 1/Q$, the device is driven to an amplitude of $A \sim 4.5$nm. In the absence of the drive, the fluctuation of the two quadrature components is due to thermomechanical noise. As expected, the noise is random in phase and therefore has a circularly symmetric distribution [23], as illustrated for clarity in a zoom-in in Fig. 2(b). Fig. 2(c) shows that the fluctuation is Gaussian distributed and has a standard deviation $\sigma$ agrees with the expected value $\sqrt{\overline{\Delta x^2}} = \sqrt{kT/\omega_r^2 m}$. Here $\overline{\Delta x^2}$ is the ensemble average of the squared displacement fluctuation. In the driven case we would expect to see the same fluctuation profile if thermomechanical noise was the dominant noise source. However, this is apparently not the case. While the amount of amplitude noise is similar to that of the thermomechanical noise, there are large fluctuations along the resonance circle. Fig. 2(d) shows the measured two quadrature components under different driving intensity. It is clear that the phase angle enclosed by the extra fluctuation remains constant when the drive is increased. Therefore, this extra phase noise is independent of the driving intensity. To confirm that the two quadrature components move indeed along the resonance circle, we show in Fig. 2(e) a dataset with a longer acquisition time of 2000s. The results therefore suggest that it is the resonator frequency that is fluctuating. Here, we point out that frequency fluctuation induced by amplitude noise via mechanical nonlinear effects cannot account for the observed phenomenon. Such effect can be estimated by $\delta\omega \approx K_3 \omega_r A \delta A$, which for our case has a value of $\delta\omega = 2 \times 10^{-4}$Hz at 5K. This value is orders of magnitude smaller than the observed fluctuation. Besides, the corresponding phase fluctuation is expected to be proportional to the amplitude, which is contrary to our observation.

To understand the observed phenomenon, we theoretically analyze a driven harmonic resonator whose resonance frequency $\omega_r(t)$ fluctuates in time. Its displacement $x(t)$ follows a stochastic differential equation

$$\ddot{x}(t) + 2\gamma \dot{x}(t) + \omega_r(t)^2 x(t) = F\cos(\omega_d t)/m + f_{th}(t)/m, \qquad (1)$$

where $F\cos(\omega_d t)$ is the external driving force, and $\gamma$ is the damping rate. $f_{th}(t)$ is the corresponding thermal fluctuation force due to coupling to the external bath. It is a white force noise assumed to have a correlation time much shorter than any time scale concerned, i.e., $\langle f_{th}(t) f_{th}(t+\tau) \rangle \propto \delta(\tau)$. The Duffing nonlinear term is assumed to be negligible, as justified from above discussions. The displacement $x(t)$ can be expressed in terms of amplitude and phase as $x(t) = A(t)\cos[\omega_d t + \phi(t)]$. Here, $\phi$ is defined as the phase relative to the drive.

Before we proceed, we would like to emphasize that there are two distinct frequency concepts, namely the resonator frequency $\omega_r$ and the instantaneous frequency $\omega_i$. The former is the natural frequency at which the resonator would oscillate in the absence of drive while the latter is the apparent oscillation frequency which is defined as $\omega_i = \omega_d + d\phi/dt$. Note that a driven passive resonator should always follow the driving frequency $\omega_d$, with a shift of $d\phi/dt$. A change of resonator frequency $\omega_r$ indirectly affects the instantaneous frequency $\omega_i$ by altering the phase $\phi = \phi(\omega_r, t)$.

For a resonator with low damping rate $\gamma \ll \omega_r(t)$ driven near resonance, i.e., $\nu(t) = \omega_r(t) - \omega_d$ and $|\nu(t)| \ll \omega_d$, it can be shown that the slowly varying complex displacement $u(t) = A(t)\exp[i\phi(t)]$ satisfies the equation $\dot{u}(t) = [-\gamma + i\nu(t)]u(t) + (F + \tilde{f}_{th}(t))/2i\omega_d m$, where $\tilde{f}_{th}(t) = 2f_{th}(t)\exp(-i\omega_d t)$ [24]. Using an appropriate integrating factor, $u(t)$ can be expressed in integral form [19] as

$$u(t) = \int_{-\infty}^{t} dt' \exp\left[-\gamma(t-t') + i\int_{t'}^{t} dt'' v(t'')\right]\left(F + \tilde{f}_{th}(t')\right) / 2i\omega_d m. \tag{2}$$

We assume that in the time span concerned the phase accumulation due to the resonator frequency fluctuation is small, i.e., $\int_{t'}^{t} dt'' v(t'') \ll 1$. This condition can always be satisfied if the chosen time span is short enough. Upon integration by parts, the amplitude and phase of the displacement, given by $A(t) \approx |\text{Im}\{u(t)\}|$ and $\phi(t) \approx \text{Re}\{u(t)\}/(F/2\gamma\omega_d m) - \pi/2$, can be re-written as

$$A(t) \approx \left|\int_{-\infty}^{t} dt' \exp[-\gamma(t-t')]\left(-F - \tilde{f}_R(t') + \tilde{g}_I(t')\frac{v(t')}{\gamma}\right) / 2\omega_d m\right| \tag{3a}$$

$$\phi(t) \approx \int_{-\infty}^{t} dt' \exp[-\gamma(t-t')]\left(v(t') + \gamma\frac{\tilde{f}_I(t')}{F} + v(t')\frac{\tilde{g}_R(t')}{F}\right) - \frac{\pi}{2}, \tag{3b}$$

where $\tilde{f}_R$, $\tilde{f}_I$ and $\tilde{g}_R$, $\tilde{g}_I$ are the real and imaginary parts of $\tilde{f}_{th}$ and $\tilde{g}_{th}(t') = \int_{-\infty}^{t'} dt'' \gamma \exp[-\gamma(t'-t'')]\tilde{f}_{th}(t'')$, respectively. The first term of Eq. (3a) gives the steady state driven amplitude $A_0 = F/2\gamma\omega_d m$. In both expressions for $A(t)$ and $\phi(t)$, the second terms represent the contribution from thermal fluctuation and the third terms represent the mixing between thermal and resonator frequency fluctuation.

The first term of Eq. (3b) is the phase noise due to resonator frequency fluctuations. Note that this is the only term that is independent of the driving force while all the other terms are inversely proportional to the drive. Therefore this term will eventually dominate if the drive is sufficiently large. Our experimental data shown in Fig. 2(d) confirms that the first term dominates in the present case. Thus, the expression for the phase can be simplified as $\phi(t) \approx \int_{-\infty}^{t} \exp[-\gamma(t-t')]\nu(t')dt' - \pi/2$. It is then straightforward to show that the transfer-function between the spectral densities of $\phi(t)$ and $\nu(t)$ is given by

$$S_\phi(\omega) = S_\nu(\omega)/(\gamma^2 + \omega^2). \tag{4}$$

Note that here we did not assume any physical model of the frequency noise nor make any assumption about the correlation time scale of $\nu(t)$. Our result is thus generally valid as long as the condition $\int_{t'}^{t} dt''\nu(t'') \ll 1$ is satisfied. This condition can always be guaranteed by considering a short enough measurement time.

One important implication of Eq. (4) is that the area under the transfer function $\int_0^\infty d\omega/(\gamma^2+\omega^2) = \pi Q/\omega_0$ is proportional to Q, which means that the amount of resonator frequency noise transferred into phase noise is larger for devices with higher Q. It can be intuitively understood as the result of steeper slope in the phase-frequency curve. Another way of looking at this is to consider the instantaneous frequency spectral density $S_f(\omega) = \omega^2 S_\varphi(\omega) = \left[\omega^2/(\gamma^2+\omega^2)\right] S_\nu(\omega)$. It can be seen that the transfer function is a high pass filter: the higher the Q, the lower the roll-off frequency and more noise is allowed to pass. This is essentially the reason why we are able to quantify the frequency noise in our ultrahigh Q devices. It is generally believed that high Q always has positive impact on device performance. Here we show however that an increase in device performance based on higher Q is accompanied by higher susceptibility to the resonator's inherent frequency noise that exists.

This intrinsic resonator frequency noise imposes a detection limit for applications that rely on the measurement of resonance shift, such as mass and force gradient sensing. The minimum resolvable frequency change $\delta\omega_{min}$ can be expressed as the change of frequency that produces the same amount of phase shift as the total phase noise within the measurement bandwidth $\Delta\omega_{bw}$, or $\delta\omega_{min} = \left(\omega_r/2Q\right)\left(\int_0^{\Delta\omega_{bw}} d\omega S_\nu(\omega)/\left(\gamma^2+\omega^2\right)\right)^{1/2}$. Contrary to the frequency detection limit imposed by thermomechanical noise [25], $\delta\omega_{min}$ does not depend on the driven oscillation amplitude. Therefore, the dynamic range of the device is no longer a determinant factor for the frequency sensitivity when the phase noise of the device is dominated by the intrinsic resonator frequency noise.

To verify and apply the results obtained above, we measure the two quadrature components and compute the amplitude and phase spectral densities by periodogram using the Hanning window function. The measurement times for 5K, 78K and 296K are chosen to be 125s, 25s, and 8s, respectively, such that the condition $\int_{t'}^{t} dt'' \nu(t'') \ll 1$ is meet. The obtained spectral densities for the amplitude and phase are plotted in Fig. 3(a) and 3(b), respectively. The amplitude and phase noise background of the drive source of our lock-in amplifier [22] is well below the device noise and therefore plays negligible role in the measured spectrum. The black solid lines show the noise contribution due to thermal fluctuation (the second terms in Eq. (3a) and (3b)). The amplitude noise matches well with the thermal term but the phase noise is significantly higher than that, which agrees with the observation from Fig. 2(a). Supported by the above analysis, we attribute this extra phase noise to resonator frequency fluctuation, whose spectral density is computed by Eq. (4) and plotted in Fig. 3(c). A $1/f$ noise clearly dominates from the lowest frequency of 0.008Hz up to 5Hz. By denoting $S_\nu(f) = S_0(T)/f$, we can extract $S_0(T)$ from the resonator frequency noise spectrum. In Fig. 4 we plot $S_0(T)$ against temperature in a log-log scale. A power law fit $aT^n$ gives an exponent of $0.94 \pm 0.10$, which strongly suggests that $S_0(T)$ is linearly proportional to temperature $T$. Therefore, the resonator frequency noise spectrum has an overall $k_B T / f$ dependence.

Previously, several processes have been identified to be the sources of resonator frequency fluctuation in nanomechanical systems, such as temperature fluctuation [15], adsorption-desorption of molecules on the device surface [16,17], and molecule diffusion along the resonator [18]. However, these known processes do not give rise to the $k_B T / f$ noise spectrum we observed. We propose that this $k_B T / f$ dependent noise spectrum can be understood within the Dutta-Dimon-Horn model [26], which applies generally to fluctuations with a broad spectrum of thermally activated relaxation times. We suggest that in the present case the fluctuations originate from defect motions within the material [15]. For crystalline solids, local elastic distortion due to a point defect can be considered as an elastic dipole [27]. Depending on its orientation, the dipole produces a distortion of the local strain under a stress field and hence locally modifies the Young's modulus. The idea of elastic dipoles can also be applied to model the inherent structural disorder in amorphous solids [28]. Following Ref. [15], we consider a defect state with two possible dipole orientations separated by an energy barrier $E$ and a thermally activated reorientation time $\tau = \tau_0 \exp(E / k_B T)$. The idea is conceptually similar to the well known two-level states (TLS) model, which has successfully explained many thermal properties of amorphous solids at low-temperature [29] as well as damping of nanomechanical resonators [30]. While the TLS model describes the situation in which atoms with two energy minima are allowed to tunnel through the energy barrier quantum mechanically, here we ignore such tunneling effects and assume that the elastic dipoles reorient through thermal activation. For a tensile-stressed beam, it can be shown that such fluctuation in elasticity leads to a resonator frequency noise spectrum of $S_\nu(\omega) = \langle \nu^2 \rangle \tau / (1 + \omega^2 \tau^2)$, where $\langle \nu^2 \rangle$ is the variance of the frequency fluctuation. Following the Dutta-Dimon-Horn model [26] and assuming the distribution of activation energy $D(E)$ to be relatively flat, i.e, $(k_B T)^n d^n D(E)/dE^n \ll D(E)$, the spectra for different activation energy can be integrated and give

$$S_\nu(\omega, T) \propto \frac{k_B T}{\omega} D(\tilde{E}), \tag{5}$$

where $\tilde{E} = -k_B T \ln(\omega \tau_0)$. The result reproduces the $k_B T/f$ dependence we observe experimentally. If we assume $\tau_0$ to be $10^{-12}$s (on the order of the inverse phonon frequency [26]), the activation energy can be estimated to have a span in range of 0.01-0.8eV, which is consistent with the typical energy scale of TLS and defect motions [26,29]. Here we would like to emphasize that the defect states model is phenomenological. Its microscopic basis remains to be verified by separate experimental approaches, for example material characterization.

In conclusion, we have studied the noise characteristics of a high Q nanomechanical resonator made of stoichiometric silicon nitride. The amplitude noise can be explained by the thermal motion of mechanical resonator, while an extra phase noise is observed. We develop a method to extract the resonator frequency fluctuation and obtain a noise spectrum with $k_B T/f$ dependence. We propose that this frequency fluctuation is due to defect states with a broad spectrum of thermally-activated elastic dipole reorientation time. On the other hand, our theory also suggests that high Q devices are more susceptible to the resonator's inherent frequency fluctuation. Our result is important for understanding the noise mechanism of the high Q silicon nitride nanomechanical resonators, as well as for applying this high performance device to sensing and oscillator applications.

**Acknowledgements**. We acknowledge funding from DARPA/MTO ORCHID program through a grant from the Air Force Office of Scientific Research (AFOSR). H.X.T acknowledges support from a Packard Fellowship and a CAREER award from the National Science Foundation.

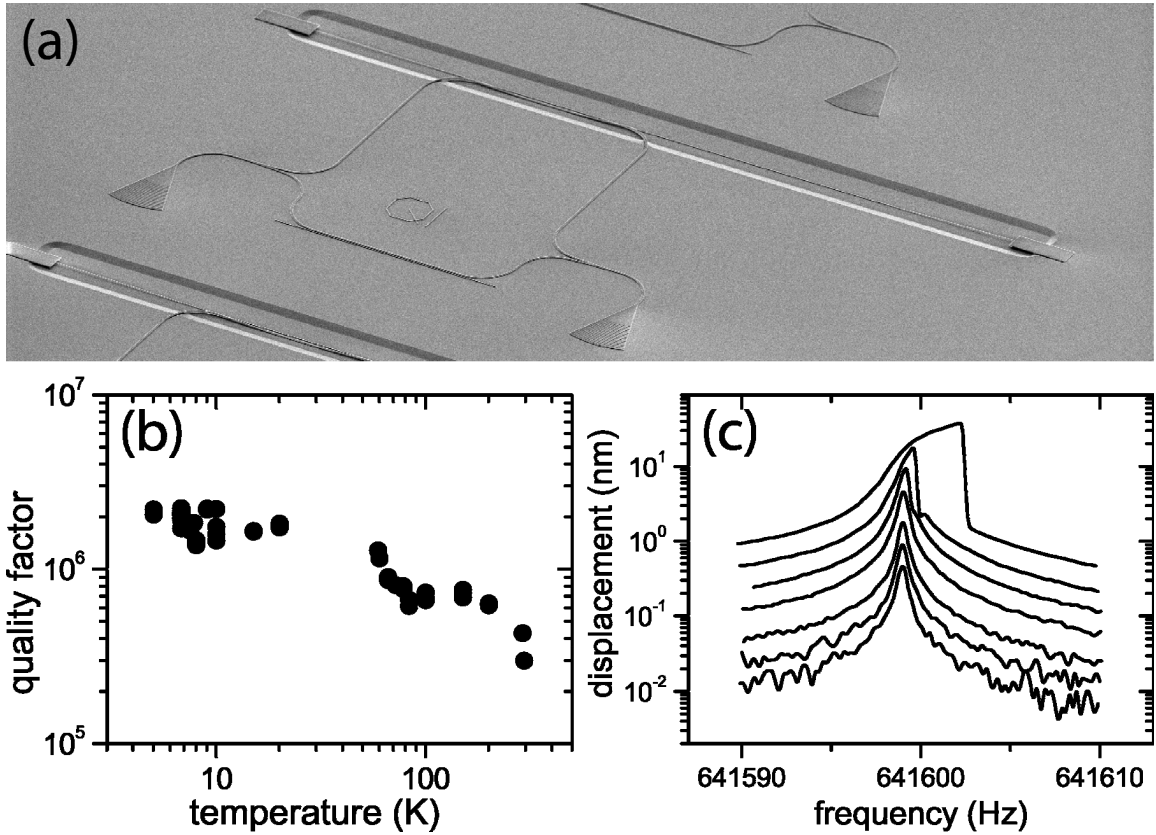

Fig. 1. (a) SEM image of the fabricated device. (b) The measured mechanical quality factor plotted against temperature. (c) The driven response of the device at different driving intensity at 5K.

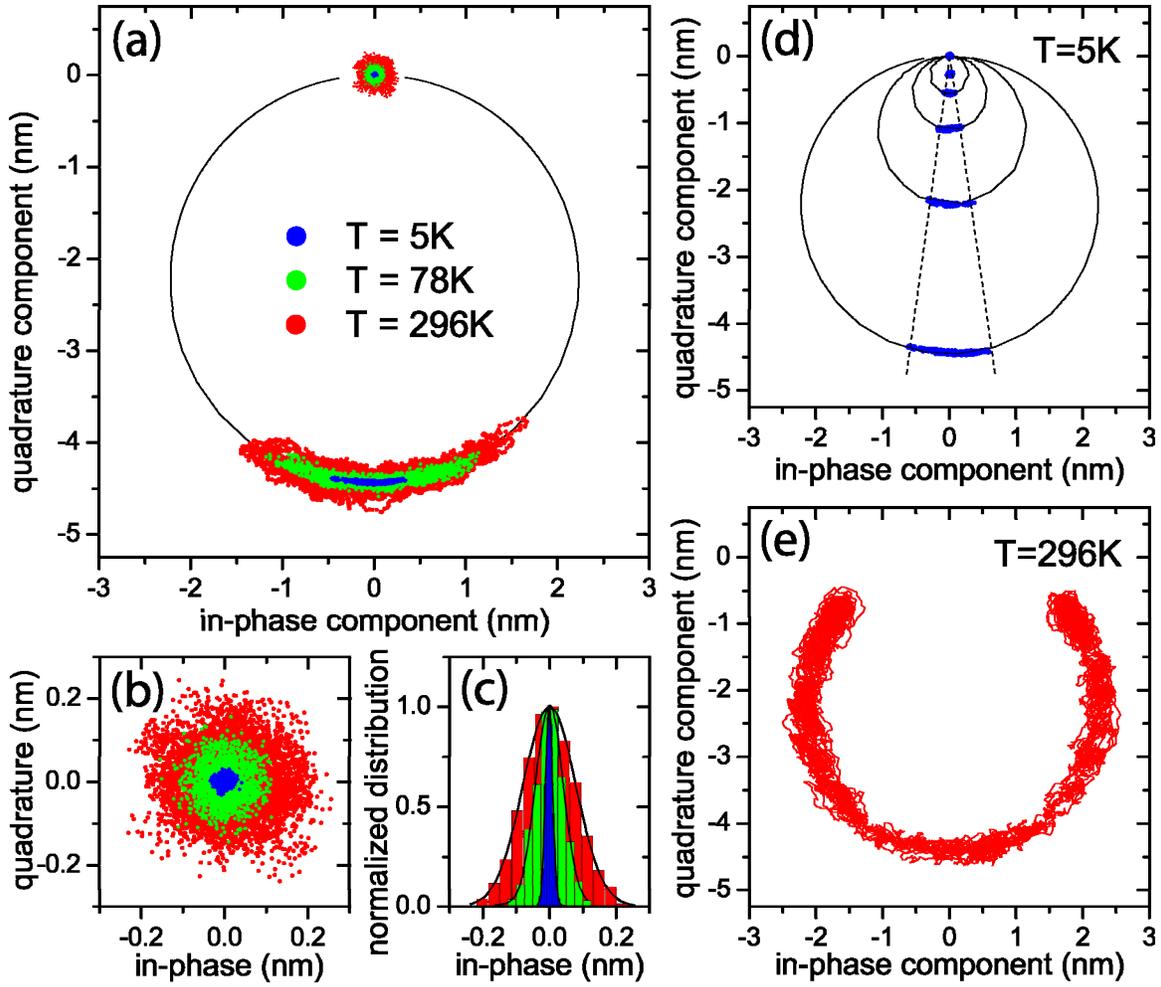

Fig. 2. (color online) (a) Two quadrature components of the device response with (lower datasets) and without drive (upper datasets) at different temperatures. The black line shows the driven response when driving frequency is swept across the resonance. (b) Zoom-in of (a) showing two quadrature components due to thermomechanical noise. (c) Normalized distribution of in-phase components shown in (b). Black lines show the Gaussian fit to the distributions. (d) Two quadrature components of the device response at different driving intensity at T=5K. Black lines show the driven response when driving frequency is swept across the resonance. (e) Two quadrature components of the device response taken in 2000s at T=296K.

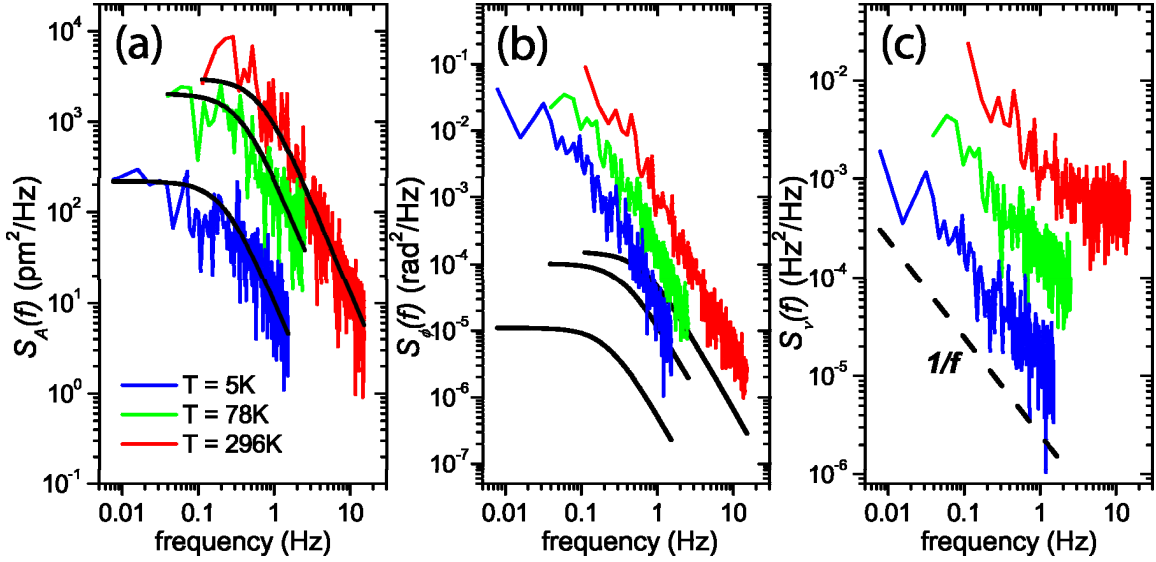

Fig. 3. (color online) Power spectral density of amplitude (a), phase (b), and resonator frequency fluctuation (c) plotted against frequency. Black solid lines show the contribution from the thermal noise given by the second terms of Eq. (3). Black dashed line shows the 1/$f$ trend.

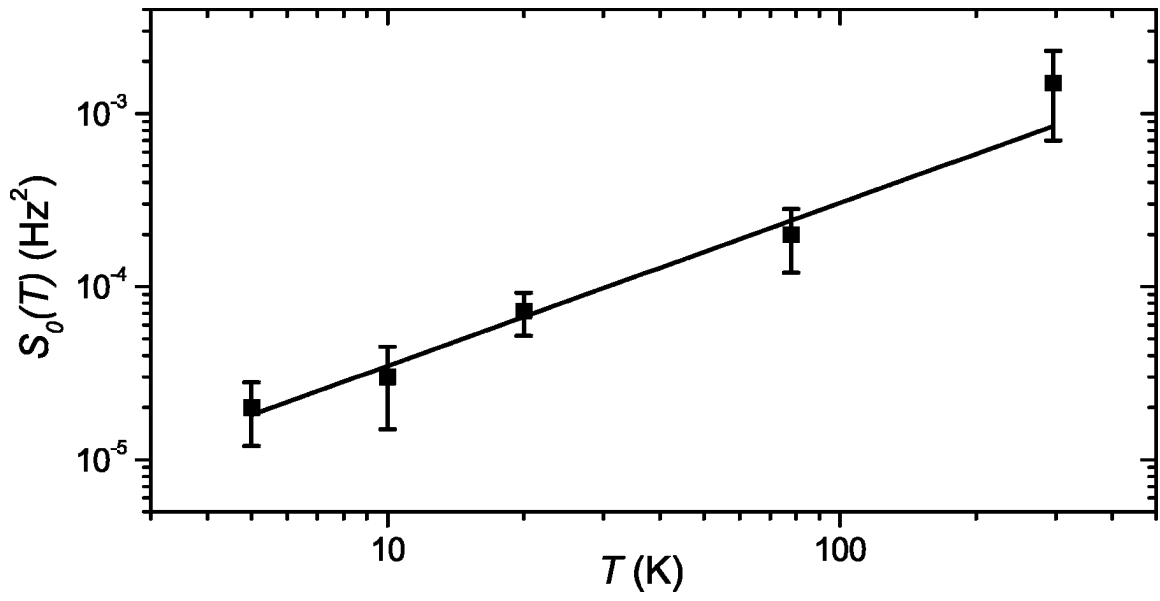

Fig. 4. Temperature dependence of $S_0(T)$. The black solid line is the power law fit $aT^n$ to the data with an exponent of $0.94 \pm 0.10$.